\documentclass{ws-ijmpd}
\usepackage{aas_macros}
\usepackage[numbers]{natbib}

\begin{document}

\markboth{Matthias Weidinger}
{Modelling the emission from blazar jets - the case of PKS 2155-304}

\title{MODELLING THE EMISSION FROM BLAZAR JETS - THE CASE OF PKS 2155-304}

\author{MATTHIAS WEIDINGER$^1$}

\address{mweidinger@astro.uni-wuerzburg.de}

\author{FELIX SPANIER$^1$}

\address{$^1$Institute for Theoretical- and Astrophysics, University of W\"urzburg\\Am Hubland, 97074 W\"urzburg, Germany}
\maketitle

\begin{history}
\received{Day Month Year}
\revised{Day Month Year}
\comby{Managing Editor}
\end{history}

\begin{abstract}
A time-dependent Synchrotron Self Compton model (SSC) which is able to motivate the used electron spectra of many SSC models as a balance of acceleration and radiative losses is introduced. Using stochastic acceleration as well as Fermi-I processes even electron spectra with a rising part can be explained, which are mandatory to fit the lowstate spectral energy distribution (SED) of PKS 2155-304 as constrained from Fermi LAT observations. Due to the time resolution the outburst of PKS 2155-304 observed by H.E.S.S. in 2006 can be modelled selfconsistently as fluctuations along the jet axis without introducing new sets of parameters. The model makes the time evolution of the SED also accessible. Hence giving new insights into the flaring behavior of blazars.
\end{abstract}

\keywords{jets; radiation mechanisms: non-thermal; BL Lacertae objects: PKS 2155-304}

\section{Introduction}
Blazars have gained a lot of attraction over the last years mainly due to new observations from imaging air Cherenkov telescopes like H.E.S.S., MAGIC or VERITAS and lately through the Fermi satellite. It is now possible to cover major parts of the very high energy (VHE) band from $30~\text{GeV}$ up to 30~\text{TeV}. The nonthermal emission of blazars is assumed to be Doppler enhanced radiation from the relativistic jet of an Active Galactic Nuclei (AGN) which axis is covering only a small angle with the line of sight. The spectral energy distribution (SED) of blazars exhibits two distinct humps. Depending on the position of those maxima blazars are divided into at least two subcategories, the highly luminous Flat Spectrum Radio Quasars (FSRQs) with the first peak occurring in the optical regime and the BL Lac Objects (high and low peaked) having the first hump up to the hard X-Rays.\\
The cause of the nonthermal radiation is not yet clarified, but a number of models have been developed and successfully applied to single objects such as hadronic models \citep{mannheim93}, Synchrotron Self Compton (SSC) and External Compton (EC) models \citep[e.g.][]{meins09, boettcher02} or the lately adopted spine layer ones \citep[e.g.][]{tavecchio08}. Unlike FSRQs BL Lac objects seem to be dominated by leptons, therefore the Compton models, where the first hump in the SED is due to synchrotron radiation and the second to upscattering of photons by electrons are very successful. We will concentrate on a SSC model where internal photons provide the targets for the inverse Compton effect. Even though these models are commonly used, there are still open questions. The first one our model will try to answer concerns the electron distribution, most models assume a fixed powerlaw rather than applying a selfconsistent evolution of electron spectra. The second one is, how the variability may be explained in this context. Standard models are not time dependent and may therefore not model variability. The model we developed is a timede-pendent SSC model, which explains the electron distribution as a balance of acceleration and radiation.\\
To test our model we choose PKS 2155-304 (henceforth shortly PKS2155) a HBL at redshift $z=0.117$ i) for it is well observed over a long period of time with the H.E.S.S. telescope giving a high confidence on the quiescent state, ii) for the recent multiwavelength campaign including Fermi \citep{fermi09} giving a good coverage of the whole SED of PKS2155. iii) PKS2155 is a very variable source with its famous flare in 2006 \citep{hess06}, the strongest outburst ever measured at these VHE. All these circumstances making it the most suitable blazar for our investigations.

\section{The Model}
In our model we solve the time-dependent kinetic equation, derived from the one dimensional diffusion approximation \citep[e.g.][]{schlickeiser84} of the Vlasov equation \citep[see e.g.][]{schlickeiser02} using the relativistic approximation $p\approx \gamma m c$, in two spatially different zones. Both zones, the acceleration zone and the radiation zone, are assumed to be spherical and homogeneous containing isotropically distributed electrons and a randomly oriented magnetic field. All calculations are made in the rest frame of the blob. Hence electrons enter the spatially smaller acceleration zone (radius $R_{\text{acc}}$) upstream the blob and are continuously accelerated due to shock and stochastic acceleration being balanced by synchrotron radiation, this extends the model of \cite{kirk98}. After $t_{\text{esc}}= \eta R_{\text{acc}}/c$ the electrons enter the radiation zone (radius $R_{\text{rad}}$) where $\eta$ is an empirical factor set to $\eta=10$. Downstream the electrons are suffering synchrotron as well as inverse Compton losses. Other contributions like pair production do not alter the SED in typical SSC conditions and are neglected \citep{boettcher02}. To calculate the model SED these photons are Doppler shifted towards the observer taking the redshift $z$ of the source into account, i.e. $I_{\nu_{\text{obs}}} = \delta^3 h\nu_{\text{obs}}/(4\pi)N_{\text{ph}}$
with $\nu_{\text{obs}}=\delta/(1+z) \nu$. Due to its size the acceleration zone does not contribute to $I_{\text{obs}}$ directly, since the observed flux at distance $r$ is given as $F_{\nu_{\text{obs}}}(r) = \pi I_{\nu_{\text{obs}}}R_{\text{rad}}^2r^{-2}$. The kinetic equation in the acceleration zone is
\begin{align}
 \label{acczone}
 \frac{\partial n_e(\gamma, t)}{\partial t} = & \frac{\partial}{\partial \gamma} \left[( \beta_s \gamma^2 - t_{\text{acc}}^{-1}\gamma ) \cdot n_e(\gamma, t) \right] + \frac{\partial}{\partial \gamma} \left[ [(a+2)t_{\text{acc}}]^{-1}\gamma^2 \frac{\partial n_e(\gamma, t)}{\partial \gamma}\right] + \nonumber \\
& + Q_0(\gamma-\gamma_0) - t_{\text{esc}}^{-1}n_e(\gamma, t)~.
\end{align}
The injected electrons, when the blob propagates through the jet, are considered via $Q_{\text{inj}}(\gamma , t) := Q_0 \delta(\gamma - \gamma_0)$. The synchrotron losses are calculated using eq. \eqref{synchrotronlosses}.
\begin{align}
 \label{synchrotronlosses}
 P_s(\gamma) & = \frac{1}{6 \pi} \frac{\sigma_{\text{T}}B^2}{mc}\gamma^2 = \beta_s \gamma^2
\end{align}
with the Thomson cross section $\sigma_{\text{T}}$. One finds $t_{\text{acc}} = \left(v_s^2/(4K_{||})+2v_A^2/(9K_{||}) \right)^{-1}$ for the characteristic acceleration timescale of the system by comparing eq. \eqref{acczone} with \cite{schlickeiser84}. $K_{||}$ is the parallel spatial diffusion coefficient which does not depend on $\gamma$ when using hard spheres. Furthermore, unlike in \citep{drury99}, the energy dependence of the escape losses is also neglected since we do not expect a pileup as suggested in \citep{schlickeiser84} at typical SSC conditions. $v_s, v_A$ are the shock and Alfv\'en speed respectively, hence $a$ in eq. \eqref{acczone} is proportional to $v_s^2/v_A^2$. All escaping electrons enter the radiation zone downstream the jet. This ansatz takes account of a much more confined shock region, for Fermi-I acceleration will probably not occur over the whole blob when considering real blazars.\\
The lack of acceleration in the radiation zone simplifies the kinetic equation to
\begin{align}
 \label{radzone1}
 \frac{\partial N_e(\gamma, t)}{\partial t}  = & \frac{\partial}{\partial \gamma}\left[\left(\beta_s \gamma^2 + P_{\text{IC}}(\gamma)\right) \cdot N_e(\gamma, t) \right] - \frac{N_e(\gamma, t)}{t_{\text{rad,esc}}} + \left(\frac{R_{\text{acc}}}{R_{\text{rad}}} \right)^3\frac{n_e(\gamma, t)}{t_{\text{esc}}}~\text{.}
\end{align}
The additional $P_{\text{IC}}$ accounts for the inverse Compton losses of the electrons occurring beside the synchrotron losses \citep[e.g.][]{schlickeiser02}:
\begin{align}
 \label{iclosses}
 P_{\text{IC}}(\gamma) & = m^3c^7h \int_{0}^{\alpha_{max}}{d\alpha \alpha \int_0^{\infty}{d\alpha_1 N_{\text{ph}}(\alpha_1) \frac{dN(\gamma,\alpha_1)}{dtd\alpha}}}
\end{align}
With the photon energies rewritten in terms of the electron rest mass, $h \nu = \alpha m c^2$ for the scattered photons and $h \nu = \alpha_1 m c^2$ for the target photons. We solve eq. \eqref{iclosses} numerically using the full Klein-Nishina cross section for a single electron \citep[see e.g.][]{jones68}, $\alpha_{max}$ accounts for the kinematic restrictions on IC scattering. Electrons escaping the blob are considered via $t_{\text{esc,rad}} = \eta R_{\text{rad}}/c$ with $\eta = 10$, which is the responding timescale of the electron system. Hence the variability time scale in the observer's frame is \citep[see e.g.][]{var95}:
\begin{align}
 \label{observersframe}
 t_{\text{var}} \propto \frac{t_{\text{esc,rad}}}{\delta}
\end{align}
To determine the time-dependent model SED of blazars the PDE for the differential photon number density, obtained from the radiative transfer equation, is solved time-dependently
\begin{align}
 \label{radzone2}
 \frac{\partial N_{\text{ph}}(\nu, t)}{\partial t} & = R_s - c \alpha_{\nu} N_{\text{ph}}(\nu, t) + R_c - \frac{N_{\text{ph}}(\nu, t)}{t_{\text{ph,esc}}}~\text{.}
\end{align}
The synchrotron production rate $R_s$ is calculated using the Melrose approximation, the inverse Compton rate $R_c$ is treated with the Klein-Nishina cross section, see \cite{meins09}. The self absorption of the synchrotron photons is described by $\alpha_{\nu}$ \citep{meins09, rueger09}. The photon loss rate is chosen to be the light crossing time.

\section{Results}
PKS2155 is used to compare the model to observations. The recently published Fermi LAT data \citep{fermi09} gives a strong constrain on the curvature of the VHE peak in the SED of PKS2155. In context of a SSC model this leads to a deep dip between the synchrotron and inverse Compton peak in the spectrum. This can only be modelled selfconsistently with a more complex electron distribution than a powerlaw used in many SSC models. We can produce an electron distribution in the acceleration zone with a rising part only by shifting the injection to moderate energies, i.e. $\gamma_0 \approx 3 \cdot 10^3$, which finally is able to explain the observed SED.
 \begin{figure}[pb]
 \centerline{\psfig{file=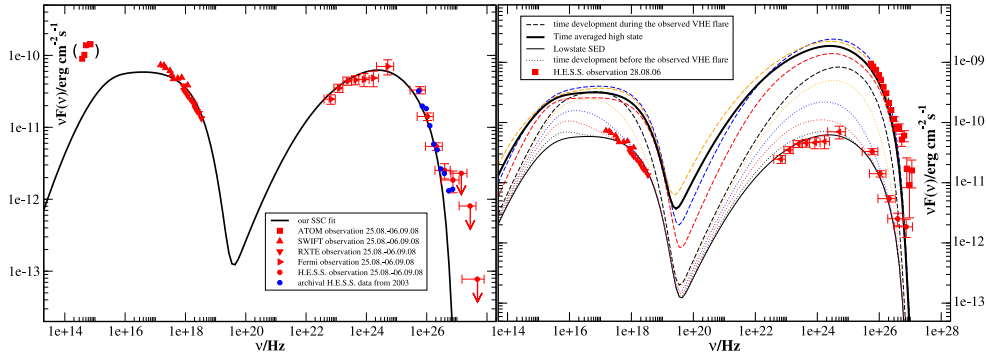,width=13.0cm}}
 \vspace*{8pt}
 \caption{a) The lowstate of PKS2155 as observed in \cite{fermi09}, red points indicate simultaneous data, blue ones are archival \citep{hess03}. The solid line represents the model SED. The ATOM datapoints have been neglegted in the modelling (see text). b) The time averaged SED over the whole flare of our model and H.E.S.S. \citep{hess06} (thick solid line) and the time evolution of the model SED during the flare.\label{f1}}
 \end{figure}
Table 1 summarizes the parameters of the model used to fit the lowstate of PKS2155 shown in Fig. 1a.
\begin{table}[ph]
\tbl{Model parameters used to fit the lowstate SED of PKS2155.}
{\begin{tabular}{@{}ccccccc@{}} \toprule
$Q_0(cm^{-3})$ & $B(G)$ & $R_{\text{acc}}(cm)$ &
$R_{\text{rad}}(cm)$ & $t_{\text{acc}}/t_{\text{esc}}$ & $a$ & $\delta$\\
 \colrule
$8.0\cdot 10^5$ & $1.4$ & $1.0 \cdot 10^{13}$ & $5.0 \cdot 10^{14}$ & $1.13$ & $20$ & $49$ \\ \botrule
\end{tabular} \label{ta1}}
\end{table}
The model SED is well in equipartition with \mbox{$h=B^2/\left(8\pi \int{\gamma N_e\text{d}\gamma mc^2} \right)=0.13$}. We neglegted the optical data assuming a contribution besides the model region (i.e. disk radiation and/or another part of the jet). However the flux level, polarization and variability indicate synchrotron radiation from the jet which is considered in other models \cite{fermi09}. This might be seen as a major drawback of our fit. This lowstate is now used to model the famous VHE outburst of PKS2155 in 2006 \citep{hess06}. In our model this is done by injecting more monoenergetic electrons at $\gamma_0$ into the acceleration zone for certain amounts of time having a reasonable physical interpretation as fluctuations in the electron density along the jet axis. These electrons also undergo acceleration and enter the radiation zone giving rise to the lightcurve for measured flux above $200~\text{GeV}$ in Fig. 2a.
 \begin{figure}[pb]
 \centerline{\psfig{file=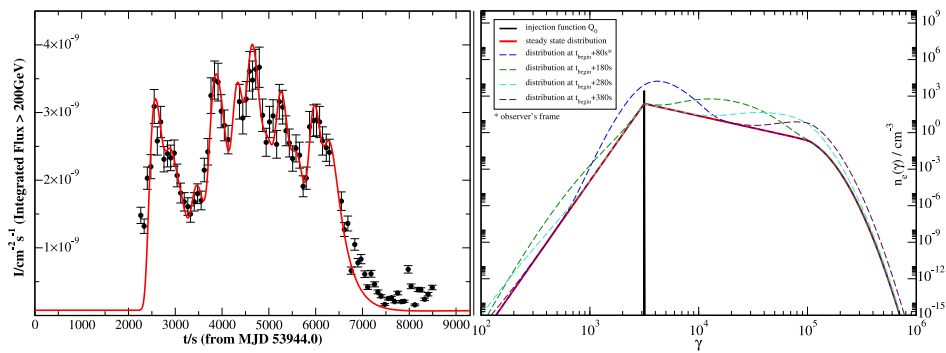,width=13.0cm}}
 \vspace*{8pt}
 \caption{a) The lightcurve of the flux above $200~\text{GeV}$ as measured by H.E.S.S. in 2006 and our model. b) Time evolution of the electrons in the acceleration zone during the first peak in the lightcurve shown in a).\label{f2}}
 \end{figure}
The variability time scale from Fig. 2a is roughly $t_{\text{var}} \approx 500~\text{s}$ when concerning the FWHM of the narrowest peak, this corresponds to the variability timescale given in eq. \eqref{observersframe} for the parameters in Table 1. Fig. 2b shows the time evolution of the electron distribution in the acceleration zone during the first peak in the lightcurve of PKS2155 out of the lowstate with its rising and falling powerlaw components and the exponential cutoff. \cite{hess06} also shows the SED of PKS2155 in the high state during the flare as a time averaged spectrum over the flare. The result of time averaging the produced model SEDs during the flare is shown in Fig. 1b. With the model however the time evolution of the SED during an outburst is also accessible, Fig. 1b. This evolution can not be observed directly due to the low flux levels from blazars in the air Cherenkov telescopes, so it is important to i) constrain the model with the available data in order to ii) get further information about the variability of blazars over the time evolution of the SED in the model.

\section{Conclusions}
The presented model is able to motivate the injected powerlaw electron spectra often used in onezone SSC models using acceleration and radiative cooling within the blob. The spectral electron index arises from the efficiency of the acceleration processes compared to the catastrophic losses in the acceleration zone. The position of the high energy cutoff is due to the equilibrium of synchrotron cooling and Fermi-I/-II acceleration, while its shape is dominated by the stochastic part of the acceleration. By injecting monoenergetic electrons at moderate $\gamma_0$, instead of low $\gamma_0$s, into the blob our model can explain electron spectra with a rising powerlaw part, as suggested in \citep{boettcher02}. Therefore the Fermi-II processes are essential. It turns out such electron spectra are mandatory to explain the curvature of the VHE peak and the deep dip between the two humps in the SED of PKS2155 as constrained by the latest Fermi observations. The moderate $\gamma_0$ factor has a physically reasonable, but highly speculative, explanation as counterstreaming electrons, since relativistic superposition yields $\gamma_0 = (1-(\Gamma \sqrt{\Gamma^2-1}+\gamma_u \sqrt{\gamma_u^2-1}(\Gamma^2+\gamma_u^2-1)^{-1})^2)^{-1/2}$, where $\Gamma$ is the Lorentzfactor of the blob and $\gamma_u$ the one of the opposed moving electrons entering the blob. Our model is able to fit the multiwavelength data, apart from the optical data, in \cite{fermi09} within equipartition ($h=0.13$) by shifting the injected electrons to $\gamma_0 = 3100$. Using this lowstate of PKS2155 as a starting point we are able to model the lightcurve of the VHE outburst measured by H.E.S.S. in 2006, even without additional parameters compared to simple onezone SSC models. Therefore we just have to inject more electrons at $\gamma_0$ into the blob for certain amounts of time thus no additional sets of parameters have to be assumed. The time average of the model lightcurve fits the H.E.S.S. time averaged SED of the flare. All this leads to a selfconsistent and highly constrained description of a HBL and the flaring process.\\
With the model we are able to plot time resolved SEDs during an outburst which are inaccessible in the VHE regime by direct observations. According to Fig 1b the changes in the X-ray regime during the flare of PKS2155 are beginning well before the H.E.S.S. observation. An average over this time would e.g. lead to a shift of the synchrotron peak or a different ratio between the two peaks. Thus one has to be carefull when comparing SEDs especially when they have been averaged over different times. To make these predictions quantitatively more exact, simultaneous observational data during an outburst is required.



\end{document}